\documentclass[twoside]{article}
\usepackage{fleqn,espcrc2}
\usepackage{graphicx,epsfig}
\newcommand{\E}[1]{$10^{#1}$eV}

\newcommand{\text}[1]{\mbox{#1}}

\title{Neutrinos and the Highest Energy Cosmic Rays}
\author{A. Letessier-Selvon\address{Laboratoire de Physique Nucl\'eaire et des Hautes \'Energies,\\
IN2P3-CNRS, Universit\'es Paris 6 \& 7,\\
4 place Jussieu, Tour 33 RdC, 75252 Paris Cedex 05 France}%
        \thanks{e-mail : Antoine.Letessier-Selvon@in2p3.fr}}
       
\begin{document}

\begin{abstract}
Observation of Ultra High Energy Cosmic Rays (UHECR) -whose energy exceeds \E{20}- 
is still a puzzle for modern astrophysics. The transfer of more than 16 Joules to a microscopic particle can 
hardly be achieved, even in the most powerful cosmic accelerators such as AGN's, GRB's or FR-II
radio galaxy lobes. Potential sources must also lie within 100 Mpc of the Earth as the interaction length of protons, 
nuclei or photons is less than 10Mpc. However no visible counterpart of those sources has been observed.
Calling upon new physics such as Topological Defect interactions or Super Massive Relic Particle  decays 
is therefore very tempting, but such objects are yet to be proven to exist. 
Due to the very low flux of UHECR only very large dedicated experiments, such as the Auger observatories,
will allow to shed some light on the origin of those cosmic rays.  In this quest neutrinos, if they can be detected, are 
an invaluable messengers of the nature of the sources.
\vspace{1pc}
\end{abstract}

\maketitle

\section{INTRODUCTION}
The cosmic ray spectrum is now proved\cite{HiresTaup99,Agasa00} to extend beyond 
\E{20}. To be observed on Earth with such energies, particles must be produced or accelerated
in the Universe  with energy near or above \E{21}. Conventional acceleration mechanisms in
astrophysical objects can only reach this requirement by stretching to the limit their available 
parameter space, making such scenarios unlikely to explain the origin of UHECR. 
Alternative hypothese involving collapse of Topological Defects (TD) or decay of Super Massive Relic Particles (SMRP)
are well suited to produce particles above \E{20} but need a proof of existence. 
\par
Transport, from the source to Earth, is also an issue. At those extreme energies the  Cosmic Microwave 
Background Radiation makes the Universe essentially opaque to protons, nuclei and photons which
suffer energy losses from pion photo production, photo-disintegration or pair production. 
These processes led Greisen, Zatsepin and Kuzmin\cite{GZK} to predict a spectral cutoff in the cosmic ray 
spectrum around $5\times$\E{19}, the GZK cutoff. The available data although still very scarce do 
not support the existence of such a cutoff. Therefore the sources are either close by and locally
over dense for the cutoff not to show, or new physics prevents the UHECR from the expected energy 
losses against the CMB photons.
\par
The following will briefly develop the arguments mentioned in this introduction. Interested readers should consult
the numerous reviews devoted to the subject\cite{Yoshida,Sigl,Bertou} for more details.

\section{OBSERVATIONS}
\begin{figure}[t]
\includegraphics*[width=7.5cm]{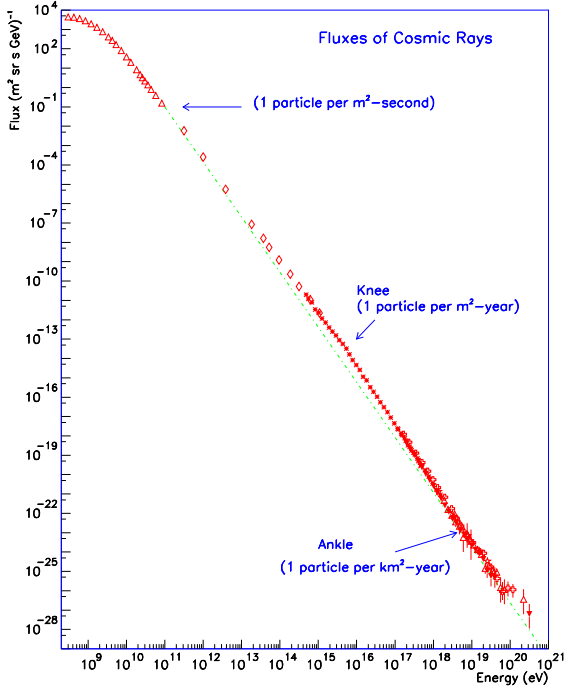}
\vspace*{-1.3cm}
\caption{The cosmic rays spectrum\cite{Swordy}}
\label{spectrum}
\end{figure}

The differential spectrum of cosmic ray flux\cite{Swordy} as a function of energy is shown on Figure~\ref{spectrum}. 
Integrated fluxes above three energy values are indicated: 1 particle/m$^2$-second
above 1~TeV, 1 particle/m$^2$-year above 10~PeV, 1 particle/km$^2$-year above 10
EeV.  The energy spectrum is surprisingly regular
in shape. From the GeV energies to the GZK cutoff, it can be represented
simply by three power-law segments interrupted by two breaks, the so-called
``knee" and ``ankle". 

\par
\begin{figure}[t]
\includegraphics*[width=7.5cm]{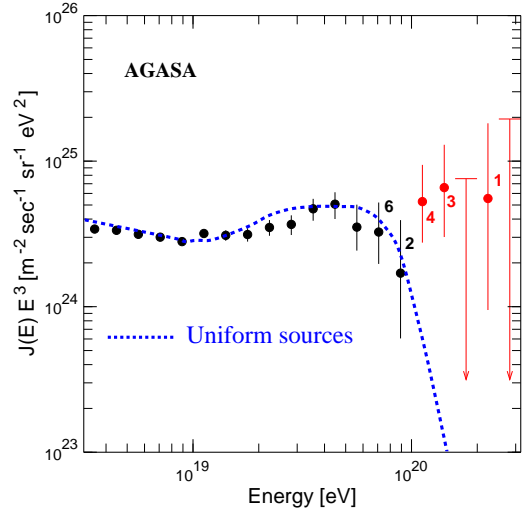}
\vspace*{-1.3cm}
\caption{Highest energy region of the cosmic ray spectrum as observed by the
AGASA\cite{Agasa00} detector.\label{takeda}}
\end{figure}

Figure~\ref{takeda} is a zoom on the
highest energy part of the total spectrum where only the
latest AGASA data\cite{Agasa00} is  displayed.
 On this figure, the energy spectrum is multiplied by $E^3$ so
that the part below the EeV energies becomes flat. Comparing the data points and the dashed line one 
 has a clear view of what can be expected from a cosmological (uniform) distribution of conventional
 sources and what is observed.
 
The cutoff, that would be expected if the sources were cosmologically and uniformly
distributed and if the observed cosmic rays had no exotic propagation or interaction
properties, \emph{is not} present in the observed data.

\par
In the search for potential sources, one
looks for correlations of the UHECR arrival directions with the distribution of matter within a
few tens of Mpc. Such an analysis 
was done by the AGASA experiment for the highest energy range~\cite{Takeda2}.
No convincing deviation from isotropy was found. 
\par
If the sources of UHECR are nearby astrophysical objects and if, as
expected, they are in small numbers,
a selection of the events with the largest magnetic rigidity would combine
into multiplets (cluster of events whose error boxes overlap).

\begin{figure}[t] 
\vspace*{-0.2cm}
\epsfig{file=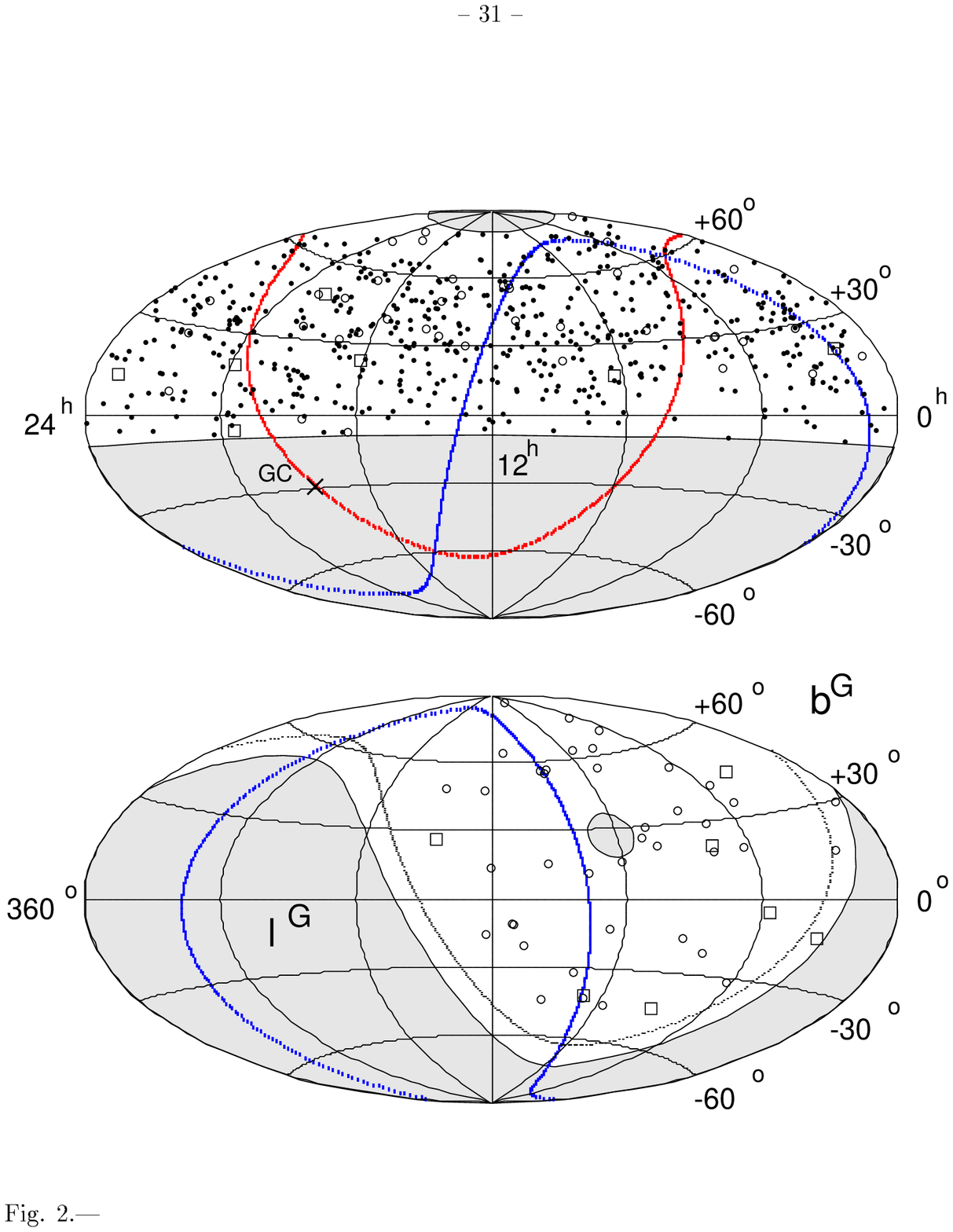,bbllx=71pt,bblly=187pt,bburx=528pt,bbury=398pt,width=7.5cm,clip=} 
\vspace*{-1cm} 
\caption{Arrival directions (galactic coordinates) of cosmic rays with $E>$40~EeV,
AGASA\label{bigevents2}.\cite{Takeda2}}
\end{figure}

Figure \ref{bigevents2} shows the subsample of events in the AGASA catalog with
energies in excess of 100~EeV (squares) and in the range 40-100~EeV (circles).
One can see that there are three doublets and one triplet. 
The chance probability of having as many multiplets as observed with
a uniform distribution is estimated to be less than 1\%~\cite{Takeda2}.

\par
The non uniform sky coverage -all present detectors are in the northern hemisphere- and the small statistics available
make anisotropy studies difficult.
The Auger observatories are designed with full sky coverage and large detection areas to overcome these  
difficulties.

\section{TRANSPORT AND PRODUCTION}
Today's understanding of the phenomena responsible for the production of UHECR 
is still limited. One distinguishes two classes of processes: the ``Top-Down''
and ``Bottom-Up'' scenarios. In Top-Down scenarios, the cosmic ray is a decay products of a 
super-massive particle. Such particles  with masses exceeding \E{21} are either meta-stable relics 
of some primordial field or a GUT gauge boson produced by the radiation or collapse of topological defects.  
In the Bottom-Up scenarios, the energy is transferred 
to a charged particle at rest through its electromagnetic interactions.
This classical approach does not require new physics.

\par
At energies above 10~EeV and except for neutrinos, the Universe 
is not transparent to ordinary stable particles on scales larger than about 100 Mpc. 
Regardless of their nature, cosmic rays lose energy in their interaction with the various photon 
backgrounds, dominantly the copious Cosmic Microwave Background (CMB) but also 
Infra-Red and Radio. The absence of prominent visible astrophysical objects in the direction of the 
observed highest  energy cosmic rays together with this distance limitation adds severe
constraints on the ``classical'' Bottom-Up picture.

\subsection{GZK cutoff}
The energy at which the Greisen-Zatsepin-Kuzmin (GZK) cutoff takes place is given by
the threshold for  pion photo-production in the proton CMB-photon collisions.
For an average CMB photon (\E{-3}), one obtains $E_{th}=7 \times $\E{19}.
The interaction length can be estimated from the pion photo-production 
cross section and the CMB density~:
\mbox{$ L=(\sigma \rho)^{-1}\simeq 6\,\mbox{Mpc}$}.       
\begin{figure}[t]
\vspace*{-0.2cm}
\includegraphics*[width=7.5cm]{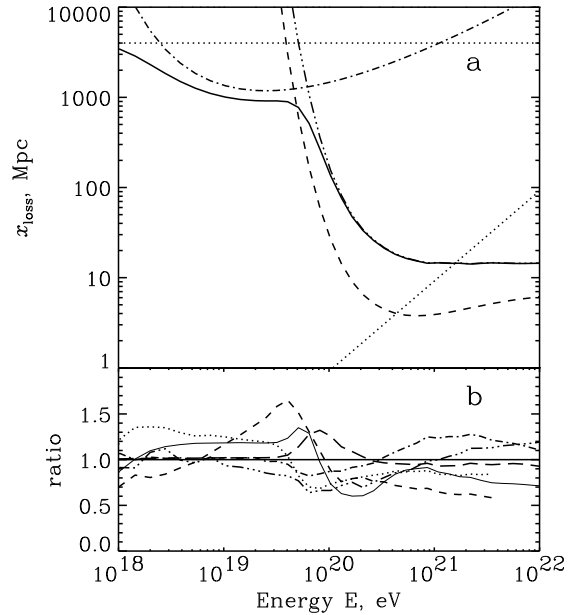}
\vspace*{-1.5cm}
\caption{Top : Loss length of a proton from~\cite{Stanev00}.
Bottom~: ratios with other calculations.}\label{GZK-loss}
\end{figure}
\par
A recent Monte Carlo~\cite{Stanev00} calculation, including red shift, pair production and pion 
photo-production losses, is shown on Figure~\ref{GZK-loss}. The loss length $x_{loss}$
is defined as  $ x_{loss}=\frac{E}{dE/dx}$. 
Above 100 EeV photo-production processes are dominant and the loss length falls below 13 Mpc. 

\par
For nuclei, the situation is usually worse. They photo-disintegrate in the 
CMB and infrared radiations losing on average 3 to 4 nucleons per Mpc when their energy 
exceeds 2$\times$\E{19} to 2$\times$\E{20} depending on the IR background density value. 
\begin{figure}[t]
\vspace*{-0.2cm}
\includegraphics*[width=7.5cm]{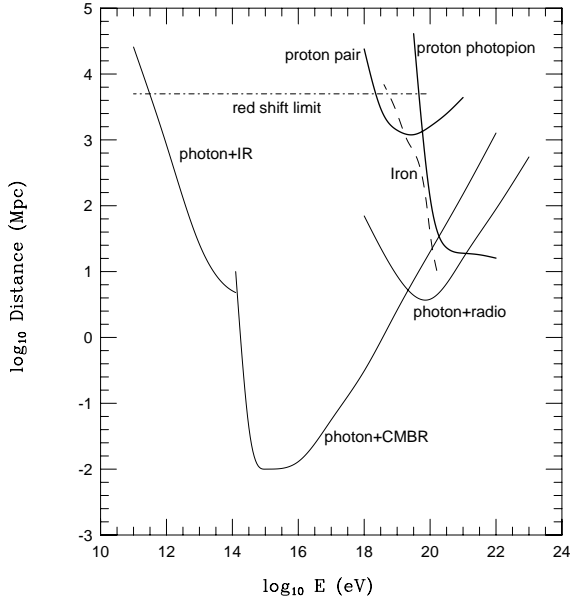}
\vspace*{-1cm}
\caption{Attenuation length of photons, protons and iron.
Double pair production (not shown) limits the photon attenuation length to about 100Mpc above \E{22}.\cite{Auger}}
\label{adrf24}
\end{figure}

\par
Top-Down production mechanisms predict that, at the source, photons and neutrinos
dominate over ordinary hadrons by about a factor four to ten\cite{Sigl,Sarkar}. 
An observed dominance of gammas or neutrinos in the supra-GZK range would then be an  
inescapable signature of a super-heavy particle decay or TD interaction. 
High energy photons traveling through the Universe produce $e^+e^-$ pairs when colliding with the 
Infra-Red/Optical (IR/O), CMB or Universal Radio Background (URB) photons. As can be seen on 
Figure~\ref{adrf24} the attenuation length gets below 10 Mpc for photon energies between $3\times$\E{13} 
and \E{20}. In this energy range the Universe is opaque to photons on cosmological scales. 

Once the photon has converted, the $e^+e^-$ pair will in turn produce photons mostly via Inverse Compton 
Scattering (ICS). Those two dominant processes  are responsible for the production of 
electromagnetic (EM) cascades.
On Figure~\ref{adrf24} one sees that, at the pair production threshold on the CMB photons 
($2\times$\E{14}), conversion occurs on distances of about 10~kpc (a thousand times smaller than for protons at 
GZK energies) while subsequent ICS of electrons on the 
CMB in the Thomson regime will occur on even smaller scales (1~kpc). 
\par
As a consequence, most photons of ultra high energy  will produce, through successive collisions
on the various photon backgrounds (URB, CMB, IR/O), lower and lower energy cascades 
and pile up in the form of a diffuse 
photon background below \E{12} with a typical  power law spectrum of index $\alpha=1.5$.  
This is a very important fact as measurements of the diffuse gamma ray background in the $10^7$-\E{11} 
range done for example by EGRET\cite{EGRET} will impose limits on the photon production fluxes of 
Top-Down mechanisms and consequently on the abundance of  topological defects or relic super-heavy 
particles.

Neutrinos are the only known particles that can travel through space unaffected even on large distances,
carrying intact the properties of the source to the observer.
They may prove to be an unambiguous signature of the new physics underlying the 
production mechanisms.

\subsection{Bottom-Up acceleration}
In the conventional acceleration scenarios one distinguishes two types
of mechanisms~:
\begin{itemize}
\item Direct acceleration by very high electric fields in or near 
very compact objects. This does not naturally provide a power-law spectrum. 
\item Diffusive shock acceleration in all systems where shock waves are present.
This statistical acceleration, known as the Fermi mechanism, 
naturally provides a power-law spectrum.
\end{itemize} 
Hillas has shown\cite{Hillas} that irrespective of the details of the acceleration 
mechanisms, the maximum energy of a particle of charge $Ze$ within a given site of size $R$ is: 
\begin{equation}
E_{\text{max}}\approx\beta Z\left(\frac{B}{1\,\mu \text{G}}\right )\left(\frac{R}{1\,\text{kpc}}\right)10^{18}~\text{eV}
\label{eq:Hillas}
\end{equation}
where $B$ is the magnetic field inside the acceleration volume and $\beta$ the velocity of the shock wave 
or the efficiency of the acceleration mechanism. This condition 
is nicely represented by the Hillas diagram shown in Figure \ref{Hillas-Diagram}.
Inspecting this diagram one sees that only a few astrophysical sources satisfy the necessary
condition given by Eq.~(\ref{eq:Hillas}). 
Let us briefly review them~:

\begin{figure}[tb]
\vspace*{-0.5cm}
\includegraphics[width=7.5cm]{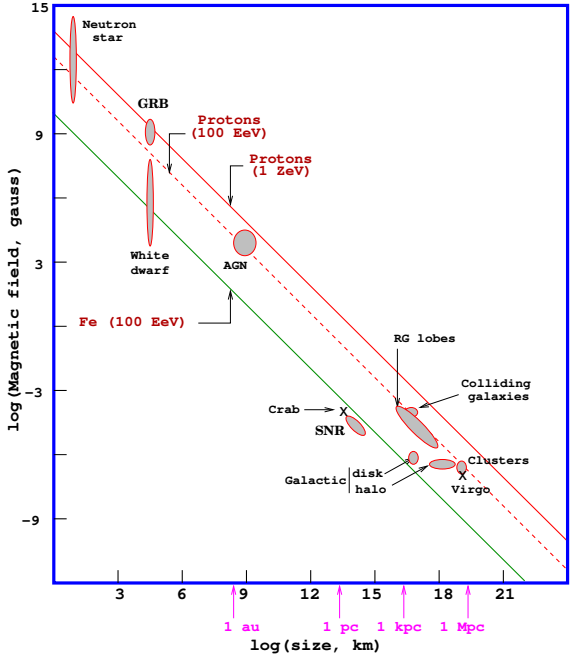}
\vspace*{-1.2cm}
\caption{Size and magnetic field strength of possible acceleration sites. Objects below the diagonal 
lines cannot accelerate the corresponding elements above \E{20} or \E{21}.\label{Hillas-Diagram}}
\end{figure}

\par{\it \bf Pulsars}~:
From a dimensional analysis, one obtains up to $10^{20}$ volts for the potential drop in a rotating magnetic 
pulsar. However the high radiation density in the vicinity of the pulsar will produce  
$e^+e^-$ pairs 
which reduce the 
potential drop down to values of about \E{13}. 
A different mechanisms involving Fe nuclei acceleration by relativistic MHD winds has been put forward~\cite{Olinto3}.
But details of the effectiveness of this mechanism still need to be demonstrated.
\par{\it \bf AGN cores and jets}~:
Blast waves in AGN jets 
could in principle lead to a maximum energy of a few tens of EeV~\cite{Zas} and similarly for AGN cores.
However those maxima are unlikely 
to be achieved under realistic conditions due to the interaction of the accelerated protons with
 very high radiation fields in and around the central engine of an AGN.

\par{\it \bf FR-II radio galaxies}~:
Radio-loud quasars are characterized by a very powerful central engine ejecting matter along 
thin extended jets. 
At the ends of those jets, the so-called hot spots, the relativistic shock wave is believed to be 
able to accelerate particles up to ZeV energies. FR-II galaxies seem the best 
potential astrophysical source of UHECR~\cite{Biermann}. Unfortunately, no nearby (less than 100~Mpc) 
object of this type is visible in the direction of the observed highest energy events. 
\par{\it \bf Gamma Ray Burst}~:
Gamma ray bursters (GRB) are intense sources of gamma rays.
The most favored GRB emission model is the ``expanding fireball model''
where one assumes that a large fireball, as it expands, becomes optically thin
hence emitting a sudden burst of gamma rays. 
The observation of afterglows allowed measurement of the red shift of the GRBs hence confirming their 
cosmological origin. 
GRB can be shown to accelerate protons up to \E{20}\cite{Waxman}. 
However in such a framework the UHECR spectrum should clearly show the GZK cut-off.

\subsection{Top-Down production}
\newcommand{\X}{$X$}
One way to overcome problems related to the acceleration of UHECR and the invisibility of their sources is to 
introduce a new unstable or meta-stable super-massive \X-particle. 
The decay of this \X-particle produces, among other things, quarks and leptons, 
 resulting in a large cascade of energetic photons, neutrinos and light leptons with a small 
fraction of protons and neutrons, part of which becomes the UHECR. 
For this scenario to be observable three conditions must be met:
\begin{itemize}
\item The decay must have occurred recently since the decay products must have traveled less than about 100~Mpc because of the attenuation processes discussed above.
\item The mass of this new particle must be well above the observed highest energy (100~EeV range), 
a hypothesis well satisfied by Grand Unification Theories (GUT) whose scale is around $10^{24}$-\E{25}.
\item The ratio of the volume density of this particle to its decay time must be 
compatible with the observed flux of UHECR. 
\end{itemize}
\noindent
The \X-particles may be produced by way of two distinct mechanisms:
\begin{itemize}
\item  Radiation, interaction or collapse of Topological Defects (TD), producing \X-particles that
 decay instantly. In those models the TD are leftovers
from the GUT symmetry-breaking phase transition in the very early universe.
Quantitative predictions of the TD density that survives a possible inflationary 
phase rely on a large number of theoretical hypotheses. Therefore they cannot be taken at face value,
although the experimental observation of large differences could certainly be interpreted as 
the signature of new effects.
\item  Super-massive meta-stable relic particles from some primordial quantum field, produced after
the now commonly accepted inflationary stage of our Universe. However the ratio of their lifetime 
to the age of the universe requires a fine tuning with their relative abundance. 
It is worth noting that in some of those scenarios the relic particles may also act as non-thermal 
Dark Matter.
\end{itemize}
In all conceivable Top-Down scenarios, photons and neutrinos dominate at the end of the hadronic cascade. 
This is \emph{the} important distinction from the conventional acceleration mechanisms.

\section{THE AUGER DETECTOR}
Large area ground based detectors 
do not observe the incident cosmic ray directly but the Extensive Air Shower (EAS), a very large cascade of particles, 
they generate in the atmosphere. All experiments aim to measure, as accurately as possible, 
the direction of the primary cosmic ray, its energy and its nature.
There are two major techniques used. One is to build a ground array of sensors spread over a large area,
to sample the EAS particle densities on the ground.
The other 
consists in studying the longitudinal development of the EAS by detecting
the fluorescence light emitted by the Nitrogen molecules which are excited by the EAS secondaries.

The Auger Observatories\footnote{Named after the French physicist Pierre Auger}~ 
combine both techniques,
with construction starting in the fall of 2000. Once completed in 2006,
they will be covering two sites,
one in the southern hemisphere (Argentina) and  one in the north
(Utah, USA). 
The surface of each site, 3000~km$^2$, will provide statistics of a few tens of
events per year above 100~EeV. The detector is designed to be fully
efficient for showers above 10~EeV, with a duty-cycle of
100\%.  Each station of the ground array is a cylindrical Cerenkov tank 
of 10~m$^2$ surface and 1.2~m height 
filled with filtered water. 
Because of the size of the array, the stations have to work in a stand-alone
mode: they are powered by solar panels and batteries, communication is wireless and timing 
is provided by the GPS satellites.

The fluorescence telescopes use photo-tubes with a field of view of $1.5^{\circ}$. 
Each telescope sees an angle of about $30\times30$
degrees. On the southern site, three eyes (7 telescopes each) will be installed
at the periphery  of the array and one (12 telescopes) in the middle, in order for
the whole array to be  visible by at least one of the telescopes. 
In the hybrid mode (10\% of the events), the detector is expected to have on
average 10\% energy resolution and an angular precision of about $0.3^{\circ}$.

With a total aperture of
14000~km$^2$sr (both sites), the Auger Observatory should detect every year of the order of
10000 events above 10~EeV and 100 above 100~EeV.

\begin{figure}[t]
\includegraphics[width=7.5cm]{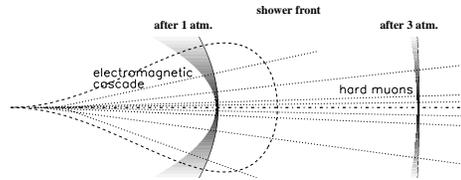}
\vspace*{-1.4cm}
\caption{Horizontal shower development~\cite{Billoir}.}
\label{showerdev}
\end{figure}

\section{NEUTRINOS}
Although both neutrinos and photons dominate the particle fluxes in Top-Down models,
only neutrinos are perfect probes of the characteristics of the sources. 
High energy neutrinos can also  be produced in Bottom Up scenarios as secondaries of hadronic interaction.
If AGNs, Radio Galaxy lobes or GRBs are UHECR sources they should  produce a substantial flux 
of neutrinos. Finally, along their path in the universe, hadrons will also produce neutrinos via the pion 
photo -production processes, the GZK neutrinos.
\par
The first study on the detection of UHE neutrinos with the Auger detector were done by \cite{Zas2,Billoir}. 
The UHE neutrinos may be detected and distinguished from ordinary hadrons by the 
shape of the horizontal EAS they produce.
At large angles, above 60$^\circ$, hadronic showers have their electromagnetic part extinguished 
and only high energy muons survive. 
Therefore the shower front is very flat (radius of curvature is larger than 100~km), 
and very narrow (less than 50~ns). Neutrinos interacting deeply in the atmosphere will start a shower above the detector 
which will appear as a ``normal'' shower, with a curved front
(a few km), a large electromagnetic component, and a wider signal (a few microseconds) [see Figure~\ref{showerdev}].
With such important differences and if the fluxes are high
enough, neutrinos will be identified and detected . 

\begin{figure}[t]
\vspace*{-0.5cm}
\includegraphics[width=7.5cm]{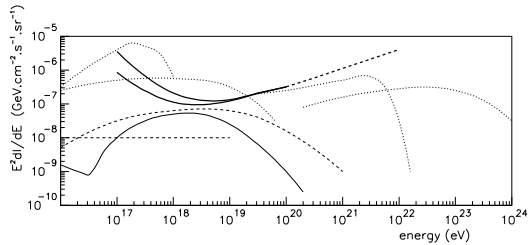}
\vspace*{-1.5cm}
\caption{Neutrino fluxes from various sources~\cite{Protheroe}, dotted lines 
speculative, dashed probable, solid certain.
The 2 top thick solid lines represent the Auger sensitivity (0.3 event per year)~\cite{Billoir}.}
\label{nu-fluxes}
\end{figure}

Figure~\ref{nu-fluxes} shows the expected fluxes from a model calculation by Protheroe~\cite{Protheroe}. 
The sensitivity limit of the Auger detector defined as 0.3 events per year is also shown. 
Although each site of the Auger observatory reaches 10 km$^3$ water equivalent of target mass, 
only the models classified as speculative by the author are expected to yield a detectable signal.

\section{CONCLUSIONS}
The composition, the shape of the energy spectrum, and the distribution of arrivals of UHECR  
will prove to be powerful tools to distinguish between the different production scenarios.

If UHECR are hadrons accelerated by Bottom-Up mechanisms, they should point 
back to their sources, with visible counterparts.

For Top-Down mechanisms and above 100 EeV, one should observe a flux of photons 
and neutrinos as the photon absorption length increases (up to a few 100 Mpc). Below 
100~EeV the spectrum shape and composition  will depend on the characteristic distance between 
TD interactions or relic particle decays and Earth, the proton attenuation length and the photon absorption 
length.  

\font\nineit=cmti9
\font\ninebf=cmbx9

\end{document}